\documentstyle[aps,preprint,epsfig]{revtex}
\begin{document}
\preprint{
\vbox{
\halign{&##\hfil\cr
	& hep-ph/0002305 \cr
	& ANL-HEP-CP-00-014\cr}}
}
\title{Top Spin and Experimental Tests\footnote{Contribution to the Thinkshop 
on Top-Quark Physics for the Tevatron Run II, Fermilab, October 16 - 18, 1999.}}
\author{Edmond~L.~Berger and Tim~M.~P.~Tait}
\address{High Energy Physics Division, Argonne National Laboratory \\
             Argonne, Illinois 60439 \\}
\date{\today}
\maketitle

\begin{abstract} 
We examine pair mass dependence near threshold as a means to measure the 
spin of the top quark in hadron collisions, and we discuss the possibility 
that a top squark signal could be hidden among the top events.  

\end{abstract} 

\section{Introduction}
\label{intro}

Evidence to date is circumstantial that the top events analyzed in 
Tevatron experiments are attributable to a spin-1/2 parent.  The 
evidence comes primarily from consistency of the distribution in 
momentum of the decay products with the pattern expected for the 
weak decay $t \rightarrow b + W$, with $W \rightarrow \l + \nu$ or 
$W \rightarrow {\rm jets}$, where the top $t$ is assumed to have 
spin-1/2.  

It is valuable to ask whether more definitive evidence for spin-1/2 
might be obtained in future experiments at the Tevatron and LHC.  We 
take one look at this question by studying the differential cross 
section $d\sigma/dM_{t {\bar t}}$ in the region near production threshold. 
Here $M_{t {\bar t}}$ is the invariant mass of the $t {\bar t}$ pair.  
We contrast the behavior of $t {\bar t}$ production with that expected 
for production of a pair of spin-0 objects.  We are motivated by the 
fact that in 
electron-positron annihilation, $e^+ + e^- \rightarrow q + {\bar q}$, 
there is a dramatic difference in energy dependence of the cross section 
in the near-threshold region for quark spin assignments of 0 and 1/2.  

For definiteness, we compare top quark $t$ and top squark $\tilde t$ 
production since a consistent phenomenology exists for top squark pair 
production, obviating the need to invent a model of scalar quark 
production.  Moreover, top squark decay may well mimic top quark decay.  
Indeed, if the chargino $\tilde \chi$ is lighter than the light top 
squark, as is true in many models of supersymmetry breaking, the dominant 
decay of the top squark is 
$\tilde t \rightarrow b + \tilde \chi^+$.  If there are no sfermions 
lighter than the chargino, the chargino decays to a W and the lightest 
neutralino $\tilde \chi^o$.  In another interesting possible decay mode, 
the chargino decays into a lepton
and slepton, $\tilde \chi^+ \rightarrow \ell^+ \tilde \nu$.
The upshot is that decays of the top squark 
may be very similar to those of the top quark, but have larger values of 
missing energy and softer momenta of the visible decay products.  
A recent study for Run~II of the Tevatron \cite{bigstop} concluded that
even with $4~{\rm fb}^{-1}$ of data at the Tevatron, and including the
LEP limits on chargino masses, these decay modes remain open (though 
constrained) for top squarks with mass close to the top quark mass .

\section{Calculation and Results}
\label{results}
At the energy of the Fermilab Tevatron, production of $t {\bar t}$ pairs 
and of $\tilde{t} {\bar{\tilde t}}$ pairs is dominated by $q {\bar q}$ 
annihilation, where the initial light quarks $q$ are constituents of the 
initial hadrons.  The subprocess proceeds through a single intermediate 
gluon at leading-order in QCD perturbation theory.  The analogy to 
$e {\bar e} \rightarrow q {\bar q}$ through an intermediate virtual 
photon is evident.  We choose to work at leading-order in the 
$t {\bar t}$ and $\tilde{t} {\bar{\tilde t}}$ partonic cross sections.

In Fig.~1(a) we display the partonic cross sections 
$\hat \sigma(\sqrt {\hat s})$ as functions of the partonic subenergy 
$\sqrt{\hat s}$ for the $q {\bar q}$ channel, where $q$ represents
a single flavor of massless quark.  We 
use the nominal value $m_t =$ 175 GeV for the mass of the top quark. 
We select $m_{\tilde t} =$ 165 GeV for the mass of the top squark so 
that the maximum values of the partonic cross sections occur at about 
the same value of $\sqrt{\hat s}$ in $t {\bar t}$ and 
$\tilde{t} {\bar{\tilde t}}$ production. Although the coupling strengths 
$g$, where $\alpha_s = g^2/(4\pi)$, are the same in the amplitudes for 
$t {\bar t}$ and $\tilde{t} {\bar{\tilde t}}$ production, the magnitude 
of the $\tilde{t} {\bar{\tilde t}}$ partonic cross section is 
a factor of $\simeq 0.015$ smaller at the peak.  The reduction comes in part 
from the final-state sum over spins and in part from the momentum dependence 
of the p-wave coupling for $\tilde{t} {\bar{\tilde t}}$ production.

If we ignore relative normalization, the very different threshold energy 
dependences of the $t {\bar t}$ and $\tilde{t} {\bar{\tilde t}}$ cross 
sections in Fig.~1(a) suggest that spin discrimination might be possible 
if one could study the pair mass distribution, 
$d{\sigma}/d M_{t {\bar t}}$.  However, in hadron reactions, one 
observes the cross section only after convolution with parton densities.  

In Fig.~1(b), we display the hadronic cross sections for 
$p {\bar p} \rightarrow t {\bar t} X$ and 
$p {\bar p} \rightarrow \tilde{t} \tilde{{\bar t}} X$ at proton-antiproton 
center-of-mass energy 2 TeV as a function of pair mass. We use the CTEQ5L
parton densities\cite{cteq5}
with the factorization scale $\mu$ chosen equal to the top quark mass for
$t \bar t$ production, and the top squark mass for $\tilde{t} {\bar{\tilde t}}$ 
production.  We include the relatively 
small contributions from the glue-glue initial state.  The parton 
luminosities fall steeply with subenergy so the tails at high pair mass 
evident in Fig.~1(a) are cut-off sharply in Fig.~1(b).  Indeed, the 
convolution with parton densities sharpens the peak of the 
$\tilde{t} {\bar{\tilde t}}$ pair mass distribution significantly and 
makes it resemble a background that is similar to $t {\bar t}$ production.
The top squark cross section is approximately 12\% of the top
cross section.  The smaller value is due in part to the fact that the 
$p-$wave top squark production reduces the partonic cross section for low 
$M_{t {\bar t}}$, where the parton luminosities are largest.

At the energy of the CERN LHC, production of $t {\bar t}$ pairs 
and of $\tilde{t} {\bar{\tilde t}}$ pairs is dominated by $g g$ 
subprocess, and the threshold behaviors in the two cases do not 
differ as much as they do for the $q {\bar q}$ incident channel.  
In Fig.~2(a), we show the partonic cross sections 
$\hat \sigma(\sqrt {\hat s})$ as functions of the partonic subenergy 
$\sqrt{\hat s}$ for the $g g$ channel. 
In Fig.~2(b), we display the hadronic cross sections for 
$p p \rightarrow t {\bar t} X$ and 
$p p \rightarrow \tilde{t} \tilde{{\bar t}} X$ at proton-proton  
center-of-mass energy 14 TeV as a function of pair mass. We include the 
relatively small contributions from the $q {\bar q}$ initial state.  
After convolution with parton densities, the shape of the 
$\tilde{t} {\bar{\tilde t}}$ pair mass distribution is remarkably 
similar to that of the $t {\bar t}$ case. 

\section{Discussion}
\label{summary}
Based on shapes and the normalization of cross sections, it is 
difficult 
to exclude the possibility that some fraction (on the order of 10\%)
of top squarks with mass close to 165 GeV is present in the
current $t {\bar t}$ sample.  The invariant mass distribution of the 
produced objects, $M_{t {\bar t}}$, is quite different at the partonic
level for the $q \bar q$ initial state
(dominant at the Tevatron), but much less so for the $g g$
initial state (dominant at the LHC).  
However, after one folds with the parton distribution functions,
the difference in the $q \bar q$ channel at the Tevatron
is reduced to such an extent that the $M_{t {\bar t}}$ distribution 
is not an effective means to isolate top squarks from top quarks.

Ironically, the good agreement of the absolute rate for 
$t {\bar t}$ production with theoretical expectations \cite{elb,mangano} 
would seem to be the best evidence now for the spin-1/2 
assignment in the current Tevatron sample.  

A promising technique to isolate a top squark with mass
close to $m_t$ would be a detailed study of the momentum 
distribution of the top quark decay products (presumably in 
the top quark rest frame).  One could look for
evidence of a chargino resonance in the missing transverse energy
and charged lepton momentum, or for unusual energy or angular
distributions of the decay products owing to the different decay chains.
One could also look for deviations from the expected correlation between
angular distributions of decay products
and the top spin \cite{parke}.  

As a concrete example of an analysis of this type, in Fig.~\ref{fig:3} 
we present the distribution in the invariant mass $X$ of the bottom quark 
and charged lepton, with $X^2 = (p_b + p_{\ell^+})^2$, where the bottom 
quark and lepton are decay products of either a top quark with 
$m_t = 175$ GeV or a top squark $\tilde t \rightarrow \tilde \chi^+ b 
\rightarrow W^+ \tilde \chi^0 b \rightarrow \ell^+ \nu_\ell
\tilde \chi^0 b$, with $m_{\tilde t} = 165$ GeV, 
$m_{\tilde \chi^+} = 130$ GeV, $m_{\tilde \chi^0} = 40$ GeV, 
and $m_b = 5$ GeV.  The $X$ distribution is a measure of the degree of 
polarization of the $W$ boson in top quark decay \cite{carlson}, and the 
figure shows that the different dynamics responsible for top squark decay 
result in a very different distribution, peaked at much lower $X$.  The 
areas under the curves are normalized to the inclusive $t \bar t$ and 
$\tilde{t} {\bar{\tilde t}}$ rates at the Tevatron and LHC, respectively.  
At the LHC there is relatively more top squark in the sample, and thus the 
difference is more prominent.

In this simple demonstration potentially important effects are ignored 
such as cuts to extract the $t \bar t$ signal from its backgrounds,
detector resolution and efficiency, and ambiguities in identifying the
correct $b$ with the corresponding charged lepton from a single decay.
Detailed simulations would be required to determine explicitly how 
effective this variable would be in extracting a top squark sample from 
top quark events. Nevertheless, such techniques,
combined with the large $t \bar t$ samples at the Tevatron
Run~II and LHC, should prove fruitful in ruling out the possibility of
a top squark with mass close to the top quark mass, or alternatively, in 
discovering a top squark hidden in the top sample.

\section*{Acknowledgments}

Work in the High Energy Physics Division at Argonne National Laboratory
is supported by the U.S. Department of Energy, Division of High Energy
Physics, under Contract W-31-109-ENG-38.  


        \begin{figure}
\vskip-0.5in
            \epsfxsize = 12 cm   
            \centerline{\epsfbox{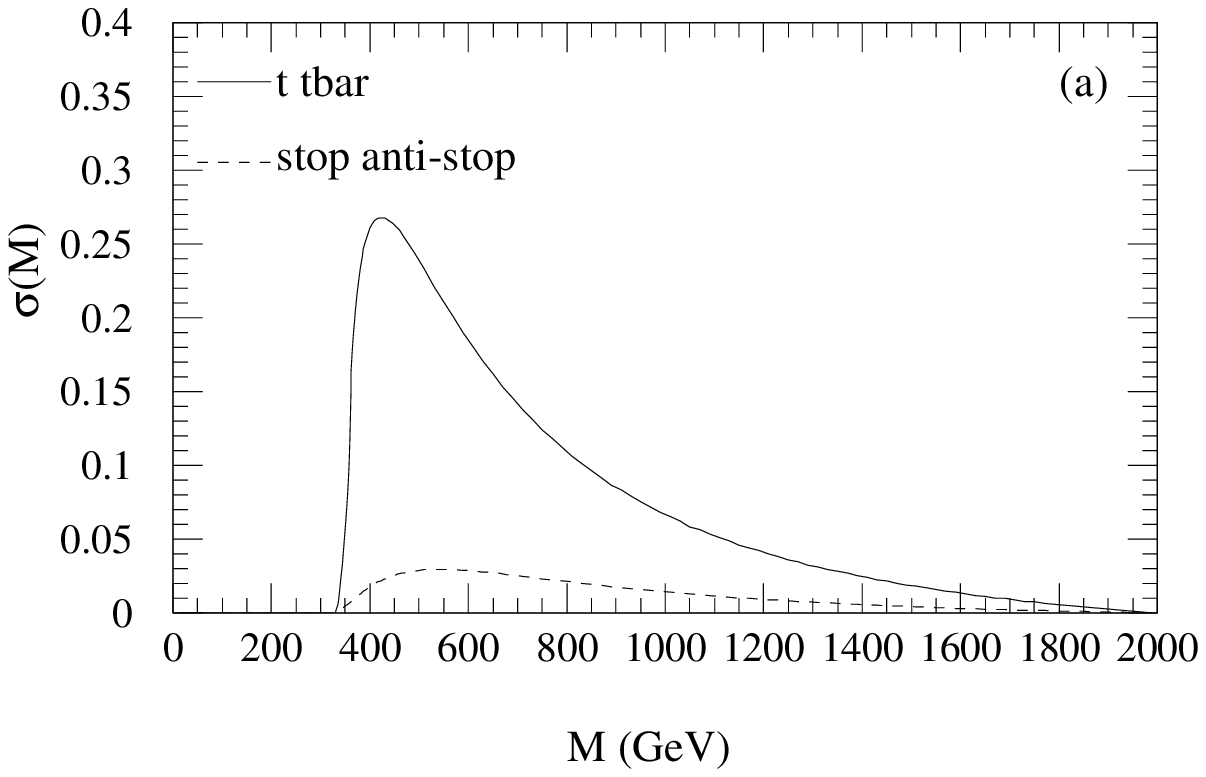}}
\vskip-0.5in          
	    \epsfxsize = 12 cm   
            \centerline{\epsfbox{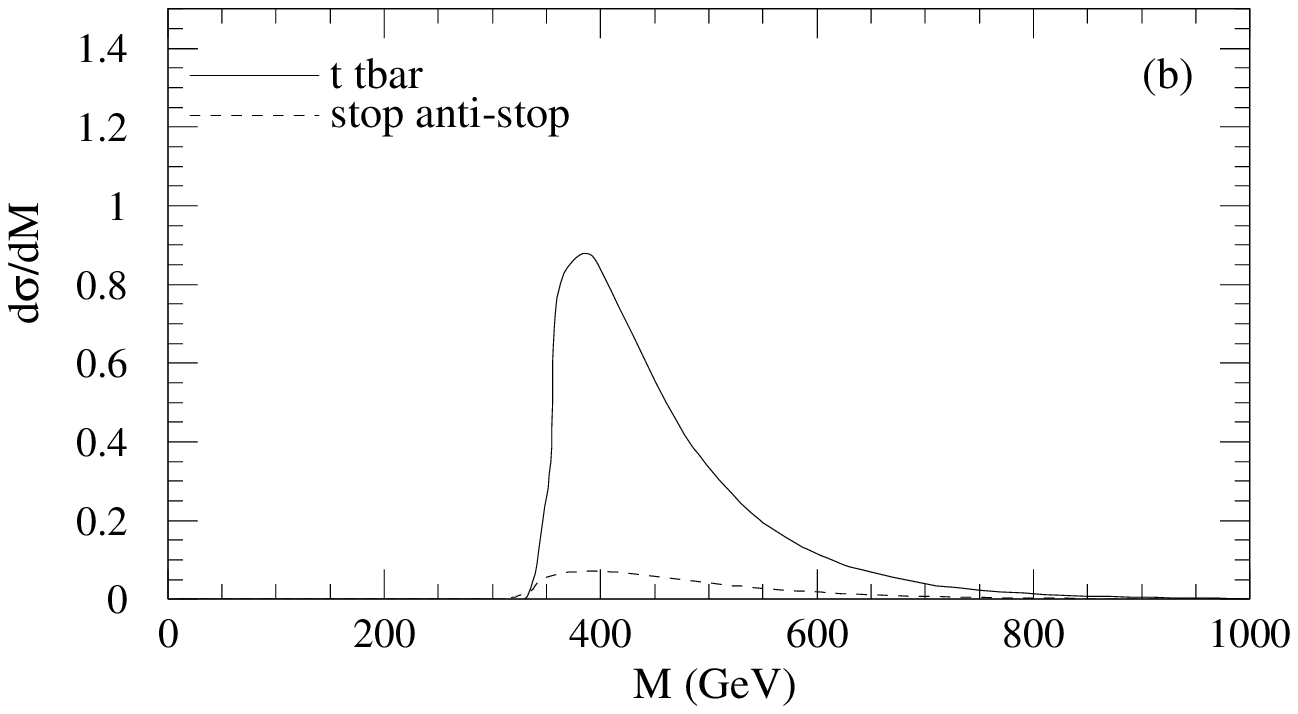}}
\vskip-0.1in
        \caption{(a) Partonic cross sections $\hat \sigma(M)$ as 
functions of partonic subenergy $M$ for the $q {\bar q}$ channel. (b) 
Hadronic cross sections $d{\sigma}/d M$ in proton-antiproton collisions 
at 2 TeV as functions of pair mass. The top quark mass $m_t = 175$ GeV, 
and the top squark (stop) mass $m_{\tilde t} = 165$ GeV.}
        \label{fig:1}
        \end{figure}

        \begin{figure}
\vskip-0.5in 
            \epsfxsize = 12 cm   
            \centerline{\epsfbox{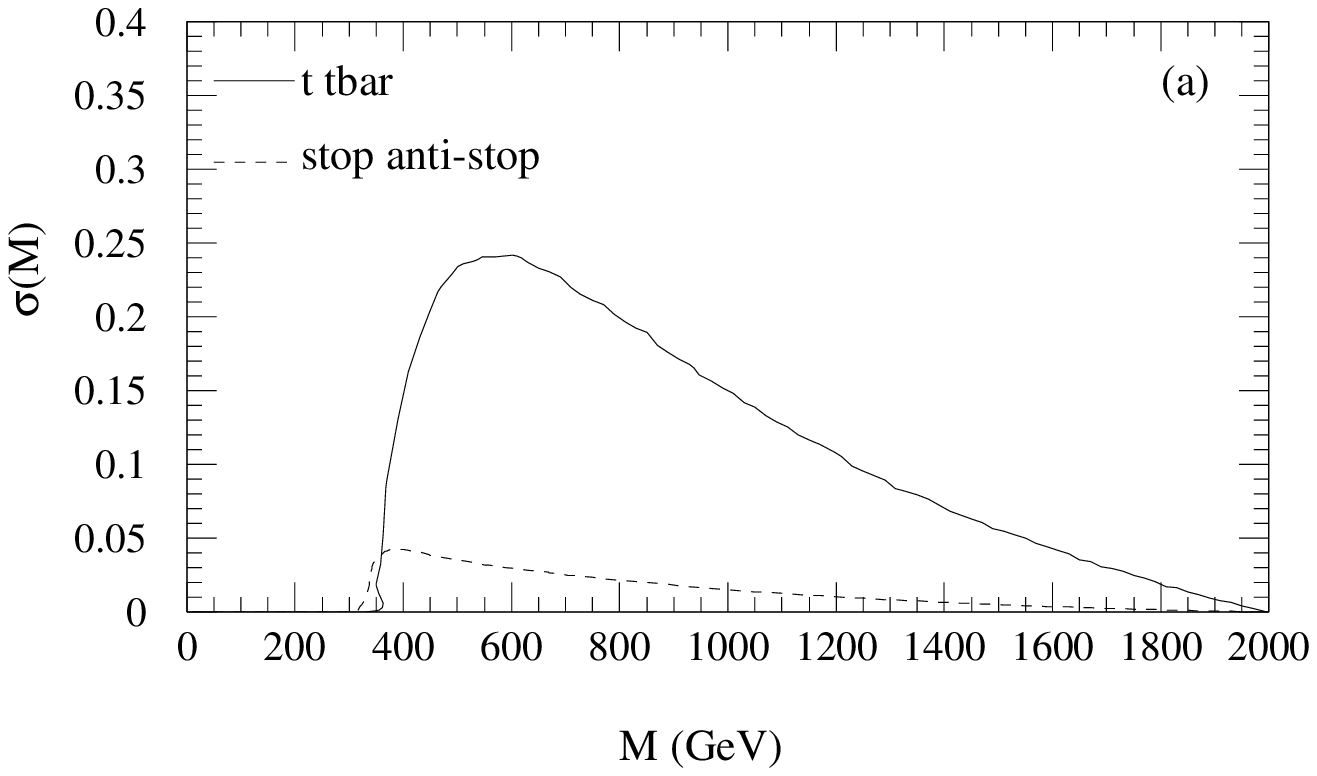}}
\vskip-0.5in 
            \epsfxsize = 12 cm   
            \centerline{\epsfbox{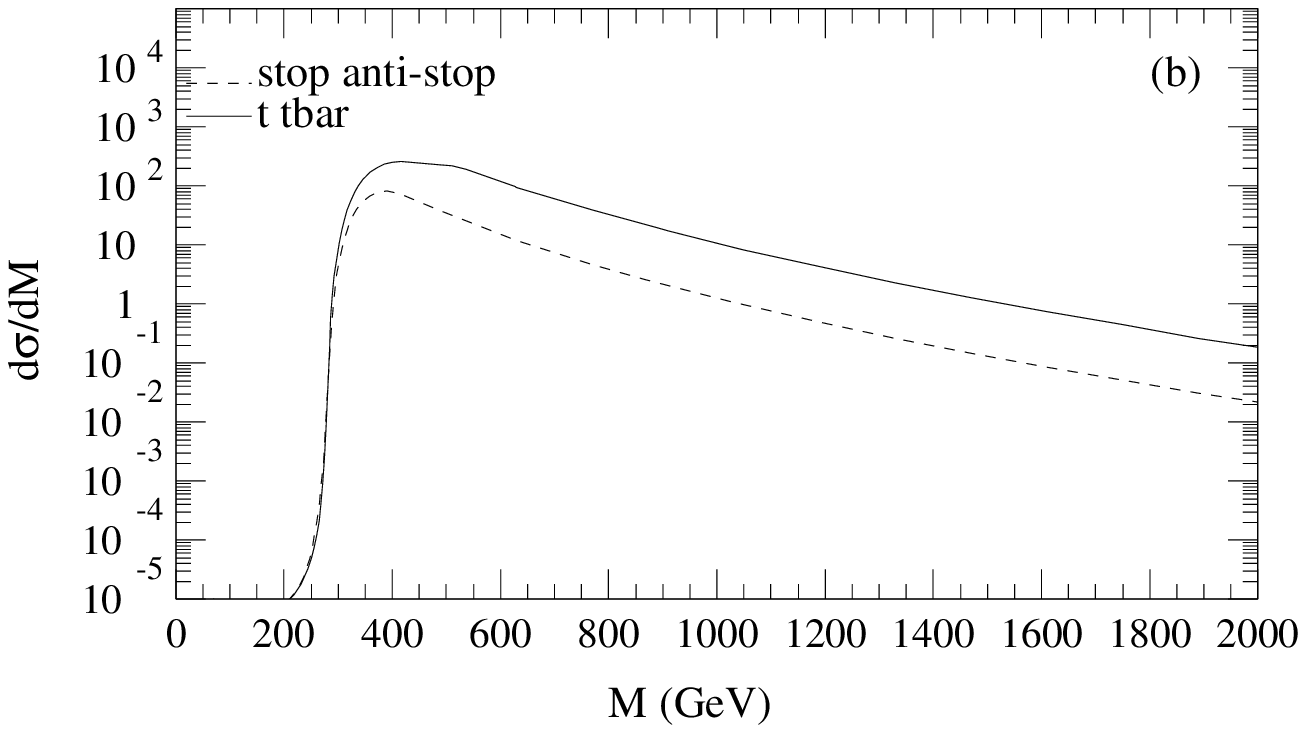}}
\vskip-0.1in 
        \caption{(a) Partonic cross sections $\hat \sigma(M)$ as 
functions of partonic subenergy $M$ for the $gg$ channel.  (b) 
Hadronic cross sections $d{\sigma}/d M$ in proton-proton collisions 
at 14 TeV as functions of pair mass. The top quark mass $m_t = 175$ GeV, 
and the top squark (stop) mass $m_{\tilde t} = 165$ GeV.}
        \label{fig:2}
        \end{figure}

        \begin{figure}
\vskip-0.5in 
            \epsfxsize = 12 cm   
            \centerline{\epsfbox{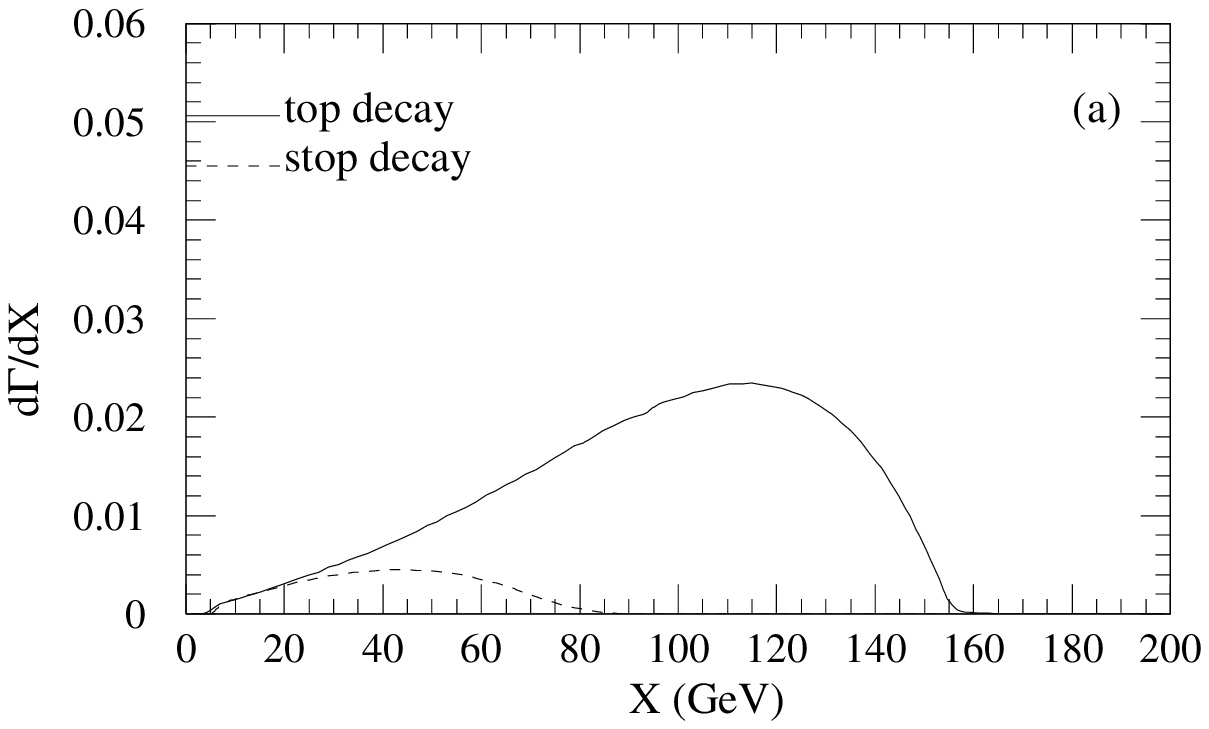}}
\vskip-0.6in 
            \epsfxsize = 12 cm   
            \centerline{\epsfbox{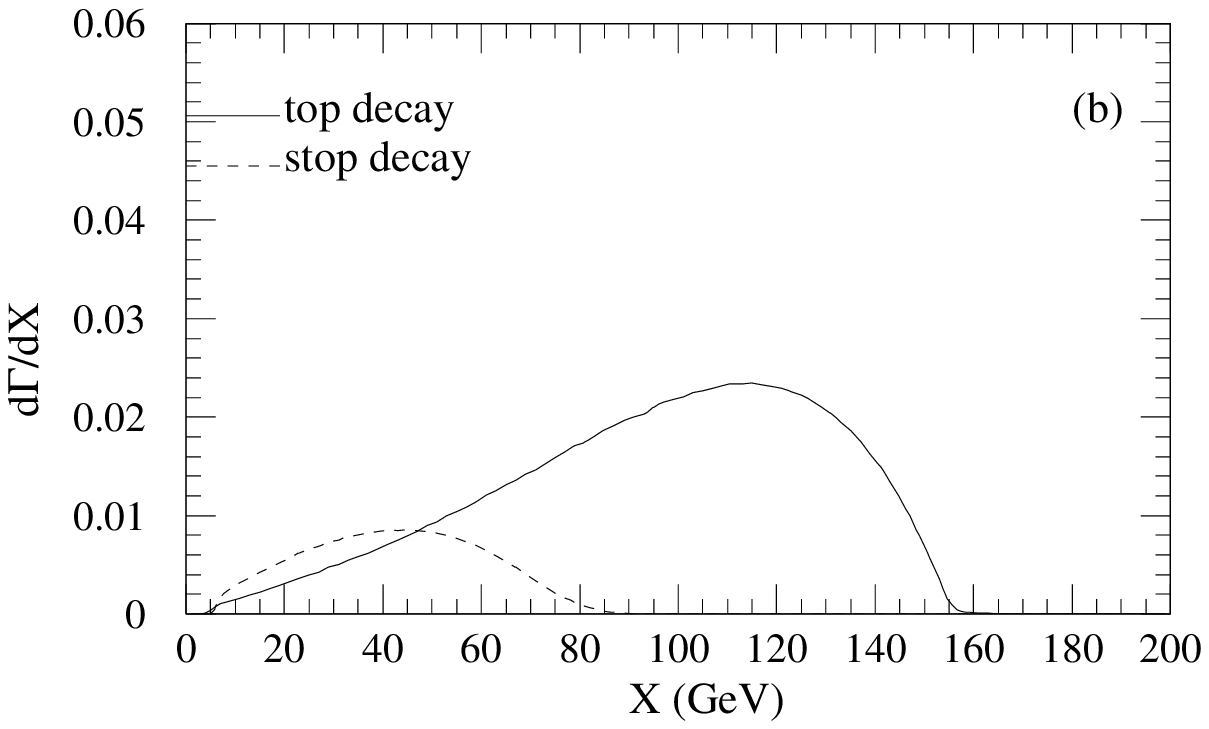}}
\vskip-0.2in 
        \caption{Distribution of the invariant mass of a bottom quark
and charged lepton ($X$) for a top quark or top squark decay, with relative 
size normalized to the cross sections at (a) the Tevatron Run~II and (b)
the LHC.  The top squark decay and sparticle masses are discussed in the 
text.}
        \label{fig:3}
        \end{figure}

\end{document}